\documentclass[%
twocolumn,
amsmath,amssymb,
aps,
prl, 10pt
]{revtex4-2}

\usepackage{graphicx}
\usepackage{dcolumn}
\usepackage{bm}
\usepackage{subcaption}

\usepackage{hyperref}

\hypersetup{
	colorlinks=true,
	linkcolor=blue,
	urlcolor=magenta,
	citecolor=red,
}

\usepackage{float}
\allowdisplaybreaks
\def\be{\begin{equation}}
	\def\ee{\end{equation}}

\begin{document}
	
	\preprint{APS/123-QED}
	
	\title{ Field theory expansions of string theory amplitudes}
	
	\author{Arnab Priya Saha}
	\email{arnabsaha@iisc.ac.in}
	\author{Aninda Sinha}%
	\email{asinha@iisc.ac.in}
	\affiliation{%
		Centre for High Energy Physics,\\
		Indian Institute of Science,\\
		C.V. Raman Avenue, Bangalore 560012, India.
	}%

	\date{\today}
	
\begin{abstract}
Motivated by quantum field theory (QFT) considerations, we present new representations of the Euler-Beta function and tree-level string theory amplitudes using a new two-channel, local, crossing symmetric dispersion relation. Unlike standard series representations, the new ones are analytic everywhere except at the poles, sum over poles in all channels and include contact interactions, in the spirit of QFT. This enables us to consider mass-level truncation, which preserves all the features of the original amplitudes. By starting with such expansions for generalized Euler-Beta functions and demanding QFT like features, we single out the open superstring amplitude. We demonstrate the difficulty in deforming away from the string amplitude and show that a class of such deformations can be potentially interesting when there is level truncation. Our considerations also lead to new QFT-inspired, parametric representations of the Zeta function and $\pi$, which show fast convergence. 
	
\end{abstract}

\maketitle	
{\bf{Introduction}}: It is often said that string theory amplitudes are very different from quantum field theory (QFT) amplitudes \cite{Green:1987sp, cappelli2012birth}. For instance, they exhibit Regge behaviour and exponentially soft high energy behaviour at fixed angle scattering. Further, it is also commonly held that string theory amplitudes, for instance for 2-2 scattering with two channels, can only be expanded in the poles of one channel but not all, as the sum over poles in one channel would automatically produce the poles in the other channel. This would give the impression, that we cannot write down a field theory representation of such amplitudes, where we have cubic vertices for the higher spin massive exchanged particles and (potential) contact diagrams. To be precise, we would like to truncate such an expansion up to some mass level. To capture the essential features of the string amplitude, like the exponentially soft high energy behaviour or the Regge behaviour up to some energy scale, if we had to retain a very  large number of mass levels, much bigger than this energy scale, we would not call this field theory. We already know that there exists a perfectly valid and well understood low energy limit, where the energy scale is much smaller than the inverse string length, and where the description is via a low energy effective action in QFT. Further, string field theory hints at the existence of a field theory like representation of string amplitudes, not just at low energies but including the higher mass poles. In this letter, we demonstrate that there indeed exists such a field theory representation of 2-2 string amplitudes, which enables us to truncate in mass-levels, while retaining all the properties of the string amplitudes.

\textcolor{black}{Besides searching for a QFT-like representation for stringy physics, a key motivation behind our investigations is the modern incarnation of the S-matrix bootstrap, which uses a crossing symmetric basis, and for numerical investigations truncates in spin and energy \cite{Paulos:2017fhb}. There are no existing representations of tree-level string amplitudes which are manifestly crossing symmetric, and admit truncation in spin and the mass-levels while capturing all the key stringy features up to some energy. While this is not an obstacle to investigate low energy constraints \cite{Guerrieri:2021ivu}, where the familiar crossing symmetric low energy expansion can be exploited, it is important to know if similar numerical bootstrap techniques have the potential to capture higher mass-level, stringy physics. Our findings in this paper point at the affirmative.}

{\bf {Background}:} The simplest string amplitude is related to the Beta function, introduced by Euler in the 18th century, which has the integral representation:

\be
B(-s_1,-s_2)=\frac{\Gamma(-s_1)\Gamma(-s_2)}{\Gamma(-s_1-s_2)}=\int_{0}^1 dz \,z^{-s_1-1}(1-z)^{-s_2-1}\,
\ee
and connects up with a string world-sheet picture.
The convergence of the integral requires ${\rm Re}(s_1)<0$ and ${\rm Re}(s_2)<0$. For other $s_1,s_2$ it is defined via analytic continuation and is given by the Gamma function form. It is well-known that the Beta function can be expanded in terms of the $s_1$ or $s_2$ poles or mass-levels:
\be\label{rep1}
B(-s_1,-s_2)=-\frac{1}{s_1}-\sum_{n=1}^\infty \frac{1}{n!} \frac{(s_2+1)_n}{s_1-n}\,,
\ee
which converges for ${\rm Re}(s_2)<0$. Here $(a)_b=\Gamma(a+b)/\Gamma(a)$ is the Pochhammer symbol. For $b=n\geq 0$ and integer, a degree $n$ polynomial in $(a+b)$ is obtained. A similar expression exists with $s_1,s_2$ interchanged. This property was responsible for the birth of string theory via the Veneziano amplitude \cite{Green:1987sp, cappelli2012birth}, where $s_1,s_2$ are related to the usual Mandelstam variables in 2-2 scattering. Dual resonance models posit that the amplitude can be expanded in terms of the poles of either $s_1,s_2$ (channels) but not both \cite{Green:1987sp, cappelli2012birth}---purportedly distinguishing themselves from quantum field theories where we expand in terms of the poles in all channels. \textcolor{black} {For instance, in place of \eqref{rep1}, it would be incorrect to add half of each channel separately since the resulting expression would only converge when ${\rm Re}(s_{1,2})<0$ and would manifestly give the wrong residues in both channels \footnote{Other possibilities of series expansions, which manifest poles in both channels, none of which are satisfactory are discussed in the supplementary material.}.}

\textcolor{black}{{\bf QFT expectations:} For a representation for tree-level string amplitudes inside $|s_{1,2}|<R$ that is similar to QFT, we would like to impose the following reasonable {\bf criteria}: (a) the poles in all channels are explicit and these are the only singularities (b)  the important features of the string amplitude  inside $|s_{1,2}|<R$ can be captured by retaining $N\sim O(R)$ levels and (c) the truncated representation can depend on field redefinition parameters. The last point can be seen by considering the following toy problem in QFT. Consider tree-level scattering $\phi_1 \phi_2\rightarrow \phi_1\phi_2$ in a theory with two massless scalars $\phi_1,\phi_2$ and a heavy scalar $\psi$ of mass $m$, interacting via the cubic vertex $g \phi_1\phi_2\psi $. This has $s$ and $u$ channels. Now perform the field redefinition $\psi\rightarrow \psi+\lambda \phi_1 \phi_2$, which would generate new contact interactions (polynomials in momenta) proportional to $(\phi_1 \phi_2)^2$ and $(\partial_\mu \phi_1 \phi_2)(\partial^\mu \phi_1\phi_2)$. It is easy to check that, expectedly, the amplitude is invariant after this redefinition. Let us now truncate the $\psi$ mode by deleting all terms involving $\psi$--this mimics level truncation in our series expansions for the string amplitude. The scattering amplitude computed now will depend on the field redefinition parameter $\lambda$; in particular, the contact terms would depend on $\lambda$. Similar arguments can be made with higher-spin massive exchanges. Thus, we expect parametric representations of the amplitude, when we have truncation. The dependence on the parameter(s) should go away when all modes are summed over, providing another {\bf stringent check}. The nontrivial challenge is to find suitable field redefinitions which allow truncation respecting point (b) above.}

\color{black}
While string field theory numerical calculations in \cite{Sen:2019jpm} and chapter 9 in \cite{ Polchinski:1998rq} suggest a mass-level expansion of the sort described above, analytic representations do not exist in the literature  \footnote{In \cite{Giddings:1986iy}, Giddings used Witten's open string field theory to derive the Veneziano amplitude, but the equivalence was at the level of the world-sheet integrals and not in a form amenable to level truncation. This was addressed in \cite{Taylor:2002bq} numerically. To date, no analytic expressions for string amplitudes at truncated level exist. }. Even numerically, it has never been demonstrated that criterion (b) can be fulfilled with a finite set of polynomial contact terms. Using the recently revived interest in crossing symmetric dispersion relations, attempts have been made to write string amplitudes as poles over all channels \cite{Sinha:2020win, Gopakumar:2021dvg, Bhat:2023puy}. However, none of these satisfy all the criteria mentioned above  and do not allow for a convenient truncation, which would retain all the features of the string amplitude.  Our goal is to find analytic expressions whose truncations respect (a)-(c) and pass the stringent check discussed above.

 \color{black}{\bf{ Explicit representations:}}
For an amplitude $\mathcal{M}\left(s_{1},s_{2}\right)=\mathcal{M}\left(s_{2},s_{1}\right)$, we will use the local two-channel symmetric dispersion relation, which can be worked out by using \cite{Raman:2021pkf} and extending \cite{Song:2023quv}--see appendix:
	\begin{eqnarray}\label{LDR-2ch}
		&&\mathcal{M}\left(s_{1},s_{2}\right) =  \frac{1}{\pi}\int_{p_0}^{\infty}\frac{\mathrm{d}\sigma}{\sigma}\left[\frac{\sigma^{2}-y}{\sigma^{2}-x\sigma+y}\mathcal{A}\left(\sigma,\frac{y}{\sigma}\right)\right].
\end{eqnarray} 
Here $\mathcal{A}(s_1,s_2)$ is the $s_1$-discontinuity or absorptive part of the amplitude. $p_0$ denotes the first pole and $s_{1}+s_{2}=x$ and $s_{1}s_{2}=y$. The absorptive part just depends on $y$ and will yield manifestly local expressions as we will see. As explained in the appendix, locality prohibits all negative powers of $x$ and the form \eqref{LDR-2ch} assumes these so-called locality or null constraints. The use of the local crossing symmetric dispersion relation  \eqref{LDR-2ch} will lead to a series representation that has poles in all channels and will also generate contact interactions (polynomials in $x,y$), and as such is in the spirit of QFT. Furthermore, adding a linear combination of the null constraints weighted by polynomials in $x,y$ will lead to new truncated representations, similar to performing field redefinitions and then truncating, as discussed above. Finding a suitable representation from here that satisfies the criterion (b) is a very difficult optimisation problem. We will rather use a trick to find a representation that obeys criteria (b), (c). 
 
Let us consider a generalised version of the Euler-Beta function given by $ \frac{\Gamma\left(\alpha-s_{1}\right)\Gamma\left(\alpha-s_{2}\right)}{\Gamma\left(\beta-s_{1}-s_{2}\right)}$. 
The $\alpha=0, \beta=0$ case is the usual Veneziano amplitude, while $\alpha=0, \beta=1$ describes massless scattering in open superstring theory. The generalised version allows for both amplitudes. We expect simple poles at each mass level and we want a finite number of spins contributing to the discontinuity (since we want to work with a finite number of QFT cubic vertices), which means that the corresponding residue has to be a polynomial. This restricts $2\alpha-\beta=p$, where $p\geq -1$ is an integer.
Using \eqref{LDR-2ch}, this amplitude can be recast as a summation over the $s_1, s_2$ poles as follows:
%
%
%
%
\begin{widetext}
	\begin{equation}\label{gen-beta}
		\frac{\Gamma\left(\alpha-s_{1}\right)\Gamma\left(\alpha-s_{2}\right)}{\Gamma\left(\beta-s_{1}-s_{2}\right)} =\sum_{n=0}^{\infty}\frac{(-1)^{p+1}}{n!}\left(\frac{1}{s_{1}-\alpha-n}+\frac{1}{s_{2}-\alpha-n}+\frac{1}{\alpha+\lambda+n}\right)
		\left(1-\alpha-\lambda+\frac{\left(s_{1}+\lambda\right)\left(s_{2}+\lambda\right)}{\alpha+\lambda+n}\right)_{n+p}
	\end{equation}
\end{widetext}
with $2\alpha-\beta=p$. Here we have used the following symmetry of the {\it lhs} to obtain the $\lambda$-dependence on the {\it rhs}.
\begin{equation}
	s_{1}\rightarrow s_{1}+\lambda, \quad s_{2}\rightarrow s_{2}+\lambda, \quad \alpha\rightarrow \alpha+\lambda, \quad \beta\rightarrow \beta+2\lambda. \nonumber
\end{equation}
Without the benefit of $\lambda$, we would not have a QFT like representation of the amplitudes, since as one can verify with \eqref{gen-beta} for $\alpha=\beta=0$ that $\lambda\rightarrow 0$ would fail to capture the exponentially soft behaviour using a mass-level truncation. The convergence of the $n$-sum in the above equation needs $\alpha-\beta-\lambda<0$, which is independent of $s_1,s_2$, unlike the representations discussed before. Hence we have met the criteria (a)-(c). 
The familiar Euler-Beta function is obtained by choosing $\alpha=\beta=0$ (and ${\rm Re~}\lambda>0$). 
%

{\bf{Open superstring:}}
When we restrict to the situation with massless poles at the lowest level, we need to set $\alpha=0$. In superstring theory, we know that $\beta=1$. For example, this amplitude describes the scattering of gluons after dressing with a kinematic factor $t_8 {\rm Tr} F^4$, with the trace being over the gauge indices and the $t_8$ giving the contraction of the spatial indices \cite{Polchinski:1998rr}. We will now give a different perspective as to why $\beta=1$.
For a field theory representation we expect the contact terms (eg. \eqref{ccont}) after summing over levels to be finite. 
Explicitly, the contact terms take the form $(\sum_n c_0^{(n)}+x\sum_n c_1^{(n)}+\cdots) t_8 {\rm Tr} F^4$. The functions of $x,y$ get converted to derivatives acting on $F^4$. For $\beta=0$, we have $c_i^{(n)}\sim 1/n$ so that the $n$-sum does not converge while for $\beta=1$ we have $c_i^{(n)} \sim 1/n^2$ whereby the $n$-sum converges. 
More generally, the $n$-dependence of the contact terms is $n^{-1-\beta}$ so that for convergence we need $\beta>0$ and since $\beta$ is an integer $\beta\geq 1$.  The requirement of polynomial residues, as discussed above needs $\beta\leq 1$ so this means that $\beta=1$ is the unique possibility. This is precisely the open superstring amplitude! We obtained this without appealing to unitarity or a world-sheet description. In recent literature, there have been attempts to zoom in to the open string amplitude using world-sheet monodromy constraints \cite{Chiang:2023quf}.

\normalcolor
The convergence of the $n$-sum (for $\alpha=0,\beta=1$) in (\ref{gen-beta}) now tells us that ${\rm Re~}\lambda>-1$. Quite nicely, the $n=0$ term gives $1/(s_1 s_2)$, with no $\lambda$-dependence. The $\lambda$-dependence in the other levels cancels out when we sum over all poles. This is a non-trivial check of the correctness of the representation. 
Therefore, the four-point open superstring amplitude satisfies the exact relation (${\rm Re~}\lambda>-1$):
\begin{widetext}
\begin{eqnarray}\label{op-st}
	\frac{\Gamma\left(-s_{1}\right)\Gamma\left(-s_{2}\right)}{\Gamma\left(1-s_{1}-s_{2}\right)} & = & \sum_{n=0}^{\infty}\frac{1}{n!}\left(\frac{1}{s_{1}-n}+\frac{1}{s_{2}-n}+\frac{1}{\lambda+n}\right)\left(1-\lambda+\frac{\left(s_{1}+\lambda\right)\left(s_{2}+\lambda\right)}{\lambda+n}\right)_{n-1}.
\end{eqnarray}
\end{widetext}
\color{black}
We note down the first few levels (poles$+$contact) below:
%
\begin{eqnarray}\label{ccont}
	t_{0} & = & \frac{1}{s_{1}s_{2}}, \nonumber\\
	t_{1} & = & \Biggl\{\frac{1}{s_{1}-1}+\frac{1}{s_{2}-1}\Biggr\} + \frac{1}{1+\lambda}, \nonumber\\
	t_{2} & = & \frac{1}{2}\Biggl\{\frac{1+s_{2}}{s_{1}-2}+\frac{1+s_{1}}{s_{2}-2}\Biggr\} + \frac{2\lambda^{2}+3\lambda+2+2\left(\lambda+1\right)x}{2 (\lambda +2)^2},\nonumber \\
	&+&\frac{y}{2 (\lambda +2)^2}\,.
\end{eqnarray}
%
The residues over the poles are written in this particular form such that when expressed in terms of Gegenbauer polynomials in $d$ dimension, the spectrum of exchanges is manifest. Positivity of the coefficients of the Gegenbauer polynomials ensures tree-level unitarity of the amplitude \cite{Arkani-Hamed:2022gsa}. Only the scalar appears at level-1 and spin-1 appears at level-2. A combination of spins appear from level-3. Our explicit expression \eqref{op-st} suggests a suitable contact term associated with each level. 

%

\begin{figure}[H]
	\centering
	\includegraphics[scale=0.5]{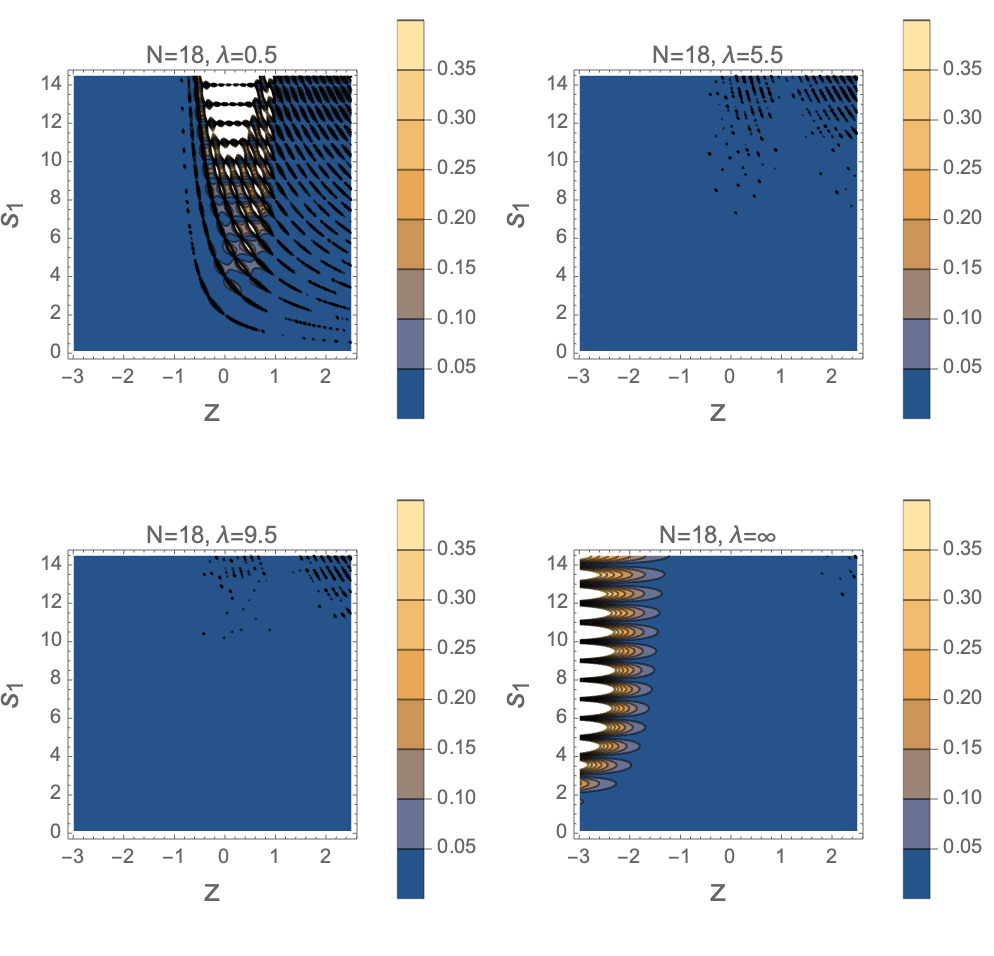}
	\caption{Fractional deviation of $M_{T}$ from the actual amplitude as function of $z=\cos\theta=1+2s_2/s_1$ for the range of $s_{1}$. $N$ denotes the number of terms retained in the truncated sum. Here $R=14$.}\label{comp1}
\end{figure}

\normalcolor
{\bf{ Efficient truncation:}} Let us now demonstrate that there exists an efficient truncation of the series representation. We will refer to the truncated sum as $M_{\text{T}}\left(s_{1},s_{2}\right)$ and number of terms in the sum by $N$. 
Let us consider energies  such that $|s_{i}|<R$. Although the sum in \eqref{op-st} contains infinite number of terms, we need to show that it is possible to truncate the sum to number of terms of $\mathcal{O}\left(R\right)$. In fig.(\ref{comp1}), we exhibit evidence that the truncated expansion, for a suitable range of $\lambda$ captures all the essential features of the actual amplitude, both in the physical region as well as in the unphysical region. 
%

\begin{figure}[H]
	\centering
	\includegraphics[scale=0.5]{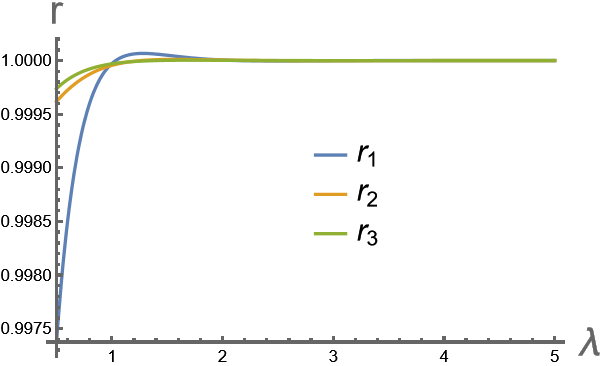}
	\caption{Ratio at each order in $s_i$  plotted as function of $\lambda$. $r_{1}$, $r_{2}$ and $r_{3}$ are the ratios at $s_i^0$, $s_i$ and $s_i^{2}$ respectively. Sum in $M_{\text{T}}$ is truncated at $N=20$.} \label{ratplot}
\end{figure}

It is also clear that various energy regimes (low, Regge, fixed-angle scattering) are all well-captured for a range of $\lambda$. 
For instance, the low energy expansion of the amplitude is given by
\begin{eqnarray}
	\mathcal{M}\left(s_{1},s_{2}\right)&\approx& \frac{1}{s_{1}s_{2}} - \zeta\left(2\right) - \zeta\left(3\right)\left(s_{1}+s_{2}\right) \nonumber\\
	&&- \zeta\left(4\right) \left(s_{1}^{2} + \frac{1}{4}s_{1}s_{2}+s_{2}^{2}\right)+ \cdots
\end{eqnarray} 
Numerical investigations  in fig.(\ref{ratplot}) suggest that for $\lambda\gtrsim 1$ with $N=20$ terms in the truncated sum, the agreement with the actual result is more than $99\%$. The distinct plateau indicates $\partial_\lambda M_{\text{T}}\approx 0$ as we would expect since the full amplitude is independent of $\lambda$.
We can also deduce new identities for $\zeta$-functions. As an example, we present
\begin{equation}
	-\sum_{n=1}^{\infty}\frac{(2 \lambda +n) }{\lambda  n^2 n! } \left(-\frac{n \lambda }{n+\lambda}\right)_n=\zeta\left(2\right).
\end{equation}
If we set $\lambda=0$, then the summand reduces to $\frac{1}{n^{2}}$. Convergence of the sum improves dramatically when $\lambda\ne 0$. Putting $\lambda=41.5$, we get convergence to 15 decimal places retaining only 40 terms, while $\sum 1/n^2$ would need 50 million terms! These representations do not exist in the mathematics literature \cite{10.11650/twjm/1500407293}.

\normalcolor

%

{\bf{Deformations:}}
We now consider a certain deformation of the amplitude, retaining crossing symmetry and having the same spectrum. The simplest way to do this is to replace the $\frac{(s_1+\lambda)(s_2+\lambda)}{\lambda+n}$ in the Pochhammer by $\frac{(s_1+\lambda)(s_2+\lambda)}{c(\lambda+n)}$, with $c$ a deformation parameter.  This changes the residues at the poles. When we examine the unitarity of the resulting partial waves, we find a restriction on $c$ and maximum $\lambda$ as depicted in the figure below. Unitarity restricts $c\geq 1$, while $\lambda$ is also restricted from above for $c>1$, unlike for $c=1$, where the residues are independent of $\lambda$. Curiously, $D=10$ exhibits a discontinuity in the plot as $\lambda_{max}$ changes from positive to negative near $c\sim 1$. Note that $D>10$ is now allowed by tree-level unitarity.

\begin{figure}[H]
	\includegraphics[scale=0.35]{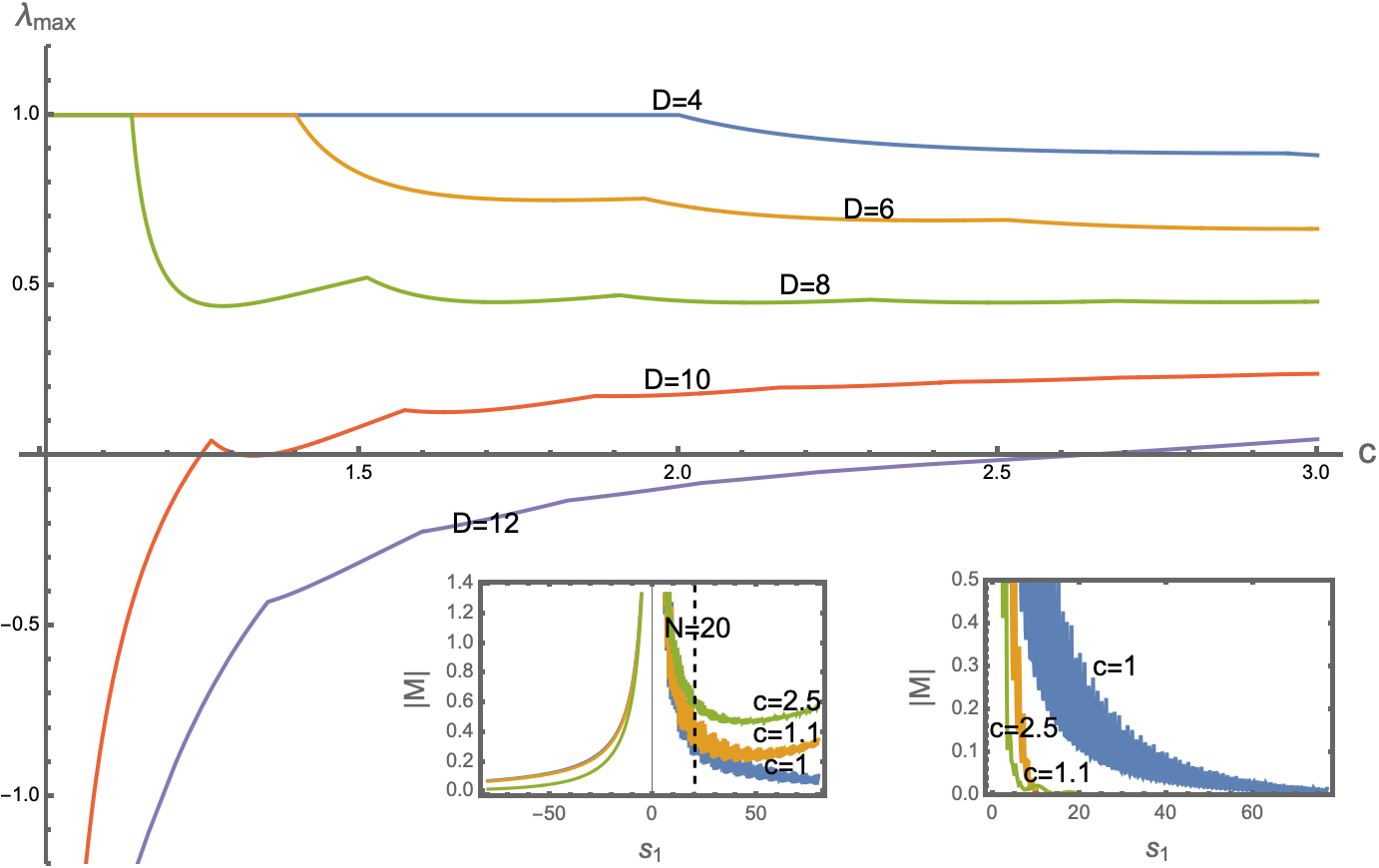}
	\caption{Tree level unitarity allows regions below the bounding lines for dimensions as indicated. Inset plots give Regge on the left and hard-scattering on the right. Here $\lambda=0.5, N=80$. }
\end{figure}

The most interesting question is if the level truncation is sensitive to the $c$-deformation or if there is some freedom which retains all the qualitative features at high energies for the amplitude. Our analysis of the high energy limits suggests that as soon as $c\neq 1$, the Regge behaviour is ruined beyond some energy scale. However, there appears to be some approximate sense in which both Regge and hard-scattering limits are similar to the stringy scenario up to some energy scale. 
We emphasise that the family of deformations we have considered does appear to single out the string theory answer, unless we restrict the energy scale.
We will leave a more detailed study of deformations  in the spirit of recent work \cite{Cheung:2023uwn, Haring:2023zwu, Chiang:2023quf} for the future. 

{\bf{Discussion:}}
There exist two well-studied limits of string theory where the string tension is either taken to infinity (eg.\cite{Green:1987sp}) or zero (eg. \cite{Gaberdiel:2021qbb}). The representations presented here can be used to study an intermediate limit, where the energy scale is such that only some of the higher spin massive modes are effectively massless. This analysis is made possible through our formulas. For instance, let us consider $M\gg |s_1|,|s_2|,\lambda\gg N$. Then we can split the sum as $\sum_{n=0}^N+\sum_{n=N+1}^M+\sum_{M+1}^\infty$. The first piece using \eqref{op-st} can be approximated and summed  to $1/(s_1 s_2 N!) (s_1+s_2+s_1 s_2/\lambda)_N$. The second piece signifies massive resonances and cannot be approximated, while the last piece can be approximated to a sum over contact interactions. The implications of these approximations will be left for future investigations. Reaching similar results using the integral formulas seems difficult to imagine.

While the analysis above has focused on the open string, similar results hold also for the closed string. {\color{black}
An explicit representation for $\mathcal{M}_{cl}(s_1,s_2)=\frac{\Gamma\left(-s_{1}\right)\Gamma\left(-s_{2}\right)\Gamma\left(-s_3\right)}{\Gamma\left(1+s_{1}\right)\Gamma\left(1+s_{2}\right)\Gamma\left(1+s_3\right)}$ is:
\begin{widetext}
\begin{eqnarray}\label{clst}
	\mathcal{M}_{cl}(s_1,s_2) =-\frac{1}{s_1 s_2 s_3}+\sum_{n=1}^{\infty}\frac{1}{\left(n!\right)^{2} }\left(\frac{1}{s_{1} -n}+\frac{1}{s_{2} -n}+\frac{1}{s_{3} -n}+ \frac{1}{n}\right)
	\left(1-\frac{n}{2} +\frac{n}{2}  \sqrt{1-4\frac{y}{n^3}}\right)^2_{n-1 }.
\end{eqnarray}
\end{widetext}
Here $y=s_1 s_2 s_3$ and $s_1+s_2+s_3=0$ and the Pochhammers yield polynomials in $y$.} A parametric representation is given in the appendix. The squared-forms of the multiplicative factors are intriguing.
Does the truncated amplitude know about the underlying worldsheet? In string field theory a dimensionless parameter enters in the plumbing fixture method, via a choice of local coordinates for fixed vertex operators on two glued discs/spheres  \cite{Polchinski:1998rq}. After Mobius transformations, a one parameter ambiguity is left. The value of this parameter controls the rate of convergence of the level sum,  similar to our findings. Our explicit representation gives an analytic realization of the expectations from string field theory. One of the most exciting prospects of the new representations in this paper is to use suitable modifications of them to re-examine experimental data for hadron scattering.

In light of our findings, it will also be interesting to re-examine \cite{Sleight:2017pcz}, which discusses a no-go theorem for quantum field theories with massless higher spin exchanges. 
Another interesting direction to pursue is to extend existing on-shell techniques \cite{Arkani-Hamed:2020blm} to incorporate contact diagrams as suggested by the analysis in this paper. In particular, when we consider higher point functions, the contact diagrams would play an important role. In principle, one could start with the absorptive part of a four-point function and obtain the contact terms using the local crossing-symmetric dispersion relation. However, in practice, this is still a difficult program to implement, since this needs to be done for amplitudes more general than just identical particles. 

Our new representation will also be useful in connecting with celestial holography \cite{Arkani-Hamed:2020gyp}. In particular, it will be interesting to examine how the mass-level truncation and the exponential softness of the full amplitude manifest in the Celestial or Carollian CFTs and compare with \cite{Stieberger:2018edy, Stieberger:2024shv}.

\section*{Acknowledgments} We thank Faizan Bhat, Rajesh Gopakumar, Prashanth Raman, Ashoke Sen, Chaoming Song, Ahmadullah Zahed, the theory group of TIFR and the participants of ISM-23, especially Sasha Zhiboedov for useful discussions. AS acknowledges support from SERB core grant CRG/2021/000873. APS is supported by DST INSPIRE Faculty Fellowship. 

\bibliography{Dispersion}

\clearpage
\onecolumngrid
\section*{APPENDIX}

\section{Crossing Symmetric Dispersion Relation}
We briefly review the two-channel crossing symmetric dispersion relation for identical scalar 2-2 scattering. 
Building on \cite{PhysRevD.6.2953}, the 2-channel symmetric dispersion relation was worked out in \cite{Raman:2021pkf} (eq 207). For $\mathcal{M}\left(s_{1},s_{2}\right)=\mathcal{M}\left(s_{2},s_{1}\right)$ and which goes like $o(s_1)$ in the Regge limit, we have
\begin{eqnarray}\label{disp-2ch-s}
		\mathcal{M}\left(s_{1},s_{2}\right) =   \mathcal{M}\left(0,0\right) 
		+  \frac{1}{\pi}\int_{a}^{\infty}\frac{\mathrm{d}\sigma}{\sigma}\mathcal{A}(\sigma,\frac{a \sigma}{\sigma-a})\left[\frac{s_{1}}{\sigma-s_{1}}+\frac{s_{2}}{\sigma-s_{2}}\right], 
\end{eqnarray}
where the $\mathcal{A}(s_1,s_2)$ is the $s_1$-discontinuity of the amplitude, $a=y/x$ with $y=s_1 s_2, x=s_1+s_2$. As it stands, this representation has non-local terms since Taylor expanding around $a=0$ leads to negative powers of $x$ which should be absent in a local theory. One can either remove these singularities imposing the so-called locality or null constraints, leading to what was dubbed as the Feynman block expansion in \cite{Sinha:2020win} or work out a more convenient local dispersion relation.
%
%
We follow a similar analysis done in \cite{Song:2023quv} for the 3-channel case to eliminate negative exponents in $x$ from the dispersion relation.
\color{black} 
First we write the kernel as $\left[\frac{s_{1}}{\sigma-s_{1}}+\frac{s_{2}}{\sigma-s_{2}}\right]=\frac{\sigma^2-y}{\sigma^{2}-x\sigma+y}-1=-1+\frac{\sigma^2-y}{\sigma^{2}+y}\sum_{n=0}^{\infty}\left(\frac{x\sigma}{\sigma^{2}+y}\right)^{n}$. We perform a Taylor series expansion of $\mathcal{A}$ around $a=0$ writing $\mathcal{A}=\sum_p c_p a^p$. The $-1$ term in the kernel will give spurious powers unless $p=0$ and hence will be replaced by $\mathcal{A}(\sigma,0)$ in the local dispersion relation. Then we have $\mathcal{A}\frac{\sigma^2-y}{\sigma^{2}-x\sigma+y}=\sum_p c_p \sum_n y^p x^{-p}\left(\frac{x\sigma}{\sigma^{2}+y}\right)^{n}\frac{\sigma^2-y}{\sigma^2+y}$  so that to avoid negative powers of $x$ we need to start the sum from $n=p$ which yields $\sum_p c_p \left(\frac{\sigma y}{\sigma^{2}+y}\right)^{p}\frac{\sigma^2-y}{\sigma^{2}-x\sigma+y}$. Therefore, the functional dependence on $a$ in $\mathcal{A}$, in \eqref{disp-2ch-s} effectively changes to $\left(\frac{\sigma y}{\sigma^{2}+y}\right)$. This leads to $\frac{a \sigma}{\sigma-a}\rightarrow \frac{y}{\sigma}$ in $\mathcal{A}$. 
We finally get \begin {eqnarray} && \mathcal
  {M}\left (s_{1},s_{2}\right ) = \mathcal {M}(0,0)+  \frac
  {1}{\pi }\int _{p_0}^{\infty }\frac {\mathrm{d}\sigma }{\sigma }\left [ \frac {\sigma ^{2}-y}{\sigma
  ^{2}-x\sigma +y}\mathcal {A}\left (\sigma,  \frac {y}{\sigma
  }\right )- \mathcal {A}(\sigma ,0)\right ]. 
  \end{eqnarray}
Here $p_0$ indicates the start of the discontinuity.
Using Cauchy's formula, the $\mathcal{M}(0,0)$ and the integral of $\mathcal{A}(\sigma,0)$ cancel for the string case, yielding in all the local dispersion relation in \eqref{LDR-2ch}. The locality constraints or null constraints are easily seen to be the sum rules arising from the difference between \eqref{disp-2ch-s} and \eqref{LDR-2ch}. Suppose we start with a series representation arising from \eqref{LDR-2ch}. To this we can always add a linear combination of the null constraints weighted by polynomials in $x,y$ which will keep the analytic structure intact (since the null constraints are just zeros). When we truncate in mass-levels,  the contact term pieces of the representation now would change and in principle this procedure can be used to find representations which admit efficient truncations \cite{Chowdhury:2021ynh}. One can use this to come up with different representations, but this seems difficult to implement in practice.

\normalcolor
\smallskip

\section{Regge $\&$ Hard Scattering}
\begin{figure}[H]
	\centering
	\begin{subfigure}[b]{0.23\textwidth}
		\includegraphics[scale=0.3]{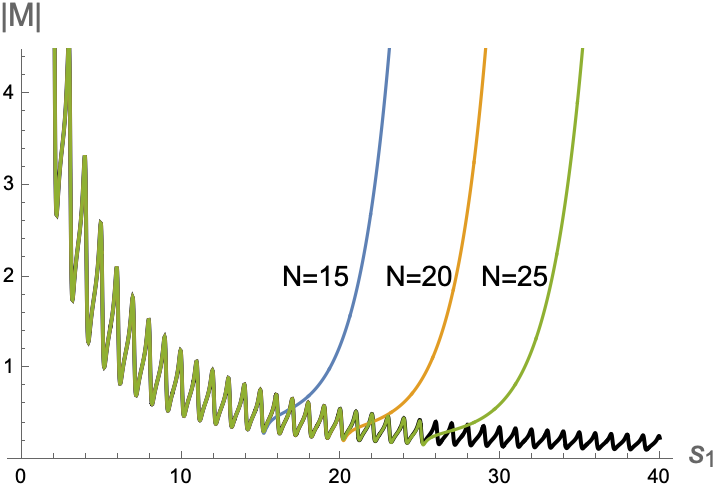}
		\caption{}\label{reggeplot}
	\end{subfigure}
	\begin{subfigure}[b]{0.23\textwidth}
		\includegraphics[scale=0.3]{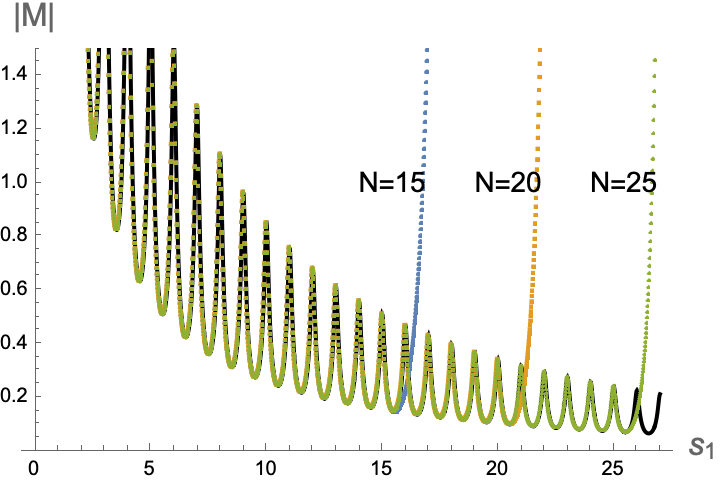}
		\caption{}\label{expplot}
	\end{subfigure}
	\caption{Here $\lambda=8.5$. The black curve represents the behavior of the actual amplitude. (a) Absolute value of truncated amplitude with $s_{2}=-0.2$ for different values of $N$ plotted as function of $s_{1}+0.1i$. (b) Absolute value of truncated amplitude with $z=-0.99$ for different values of $N$ plotted as function of $s_{1}+0.1 i$.}
\end{figure}

Let us present more evidence for the high energy limits. 
In the Regge limit $s_{1}\rightarrow\infty$ keeping $s_{2}$ fixed, the amplitude is known to behave as
$
M\left(s_{1},s_{2}\right)\approx -s_{1}^{-1+s_{2}}\Gamma\left(-s_{2}\right)\frac{\sin\left(\pi\left(s_{1}+s_{2}\right)\right)}{\sin\left(\pi s_{1}\right)}.
$
This is very accurately captured by the truncated representation as shown in fig.(\ref{reggeplot}) upto (and a bit beyond) the truncation. 
The exponentially soft high energy, fixed angle ($z=1+2s_2/s_1$) scattering behaviour is a characteristic of string theory. The truncated representation captures even this behaviour faithfully as shown in fig.(\ref{expplot}).

\normalcolor

\section{Closed-string amplitude}
We briefly discuss series representations for massless scattering in closed-string theory. In what follows $s_1+s_2+s_3=0$. 
We can work out a parametric representation of the Virasoro-Shapiro amplitude by utilizing the 2-channel symmetric local dispersion relation eq.(\ref{gen-beta}).
We start with
$\mathcal{M}_{\text{VS}}\left(s_{1}, s_{2}\right) = \frac{\Gamma\left(\alpha-s_{1}\right)\Gamma\left(\alpha-s_{2}\right)\Gamma\left(\gamma+s_{1}+s_{2}\right)}{\Gamma\left(\beta+s_{1}\right)\Gamma\left(\beta+s_{2}\right)\Gamma\left(\delta-s_{1}-s_{2}\right)}.$
Analogous to the representation of generalised Euler-Beta function of \eqref{gen-beta}, this remains invariant under $
s_{1} \rightarrow s_{1}+\lambda, \quad s_{2} \rightarrow s_{2}+\lambda,
\alpha \rightarrow \alpha+\lambda, \quad \beta \rightarrow \beta-\lambda, \quad \gamma \rightarrow \gamma-2\lambda, \quad \delta  \rightarrow \delta+2\lambda.$
Crossing symmetry in the three channels require $\gamma=\alpha$ and $\delta=\beta$. For massless pole to be present at $n=0$, we need $\alpha=0$. It can be checked that if $\beta>1$, residues in $s_{1}$ contains poles in $s_{2}$ and vice-versa, thus locality is violated. Contact terms diverge when $\beta\le0$. Thus, as in the open string case, again the superstring $\beta=1$ answer for $\alpha=0$ is singled out. Choosing $\alpha=0$ and $\beta=1$, we obtain the following one-parameter family of representations for Virasoro-Shapiro amplitude, 
%
	\begin{eqnarray}
		\frac{\Gamma\left(-s_{1}\right)\Gamma\left(-s_{2}\right)\Gamma\left(s_{1}+s_{2}\right)}{\Gamma\left(1+s_{1}\right)\Gamma\left(1+s_{2}\right)\Gamma\left(1-s_{1}-s_{2}\right)}  &=&  \sum_{n=0}^{\infty}\frac{1}{\left(n!\right)^{2}}\Biggl\{\left[\frac{1}{s_{1}-n}+\frac{1}{s_{2}-n}+\frac{1}{\lambda+n}\right] \left(1-\lambda +\frac{\left(s_{1}+\lambda\right)\left(s_{2}+\lambda\right)}{\lambda+n}\right)^2_{n-1}
		\\
		&& - \frac{1}{s_{1}+s_{2}+n}\left(1-\frac{n}{2}+\frac{\left(n-2\lambda\right)}{2}\sqrt{1-\frac{4\left(s_{1}+\lambda\right)\left(s_{2}+\lambda\right)}{\left(n-2\lambda\right)^{2}}}\right)_{n-1}^2 \Biggr\} \,. \nonumber
	\end{eqnarray}
%
The $n=0$ term corresponds to the massless pole $-\frac{1}{s_{1}s_{2}\left(s_{1}+s_{2}\right)}$. The full crossing symmetry can be made manifest level-wise by regrouping terms.

A representation for $	\frac{\Gamma\left(\alpha-s_{1}\right)\Gamma\left(\alpha-s_{2}\right)\Gamma\left(\alpha-s_{3}\right)}{\Gamma\left(\beta+s_{1}\right)\Gamma\left(\beta+s_{2}\right)\Gamma\left(\beta+s_{3}\right)}$ that starts with the fully crossing symmetric dispersion relation \cite{Sinha:2020win, Song:2023quv}   can be  delineated for $\alpha<\beta$ with the condition $2\alpha-\beta=p$ being an integer, as follows:
\begin{eqnarray}\label{genVS3}
	\sum_{n=0}^{\infty}\frac{(-1)^{n+1}}{n! \Gamma (n+\alpha +\beta )}\left(\frac{1}{s_{1}-\alpha -n}+\frac{1}{s_{2}-\alpha -n}\right.
	\left.+\frac{1}{s_{3}-\alpha -n}+ \frac{1}{\alpha
		+n}\right)\mathcal{X}_{+}\mathcal{X}_{-},
\end{eqnarray}
where $\mathcal{X}_{\pm}= \left(1-\frac{3\alpha}{2}-\frac{n}{2} \pm\frac{1}{2} (n+\alpha ) \sqrt{1-4\frac{s_{1}s_{2}s_{3}}{(n+\alpha )^3}}\right)_{n+p}$. Note that Pochhammers in the above representations always occur in pairs and are of the form, $\left(a+b\right)_{m}\left(a-b\right)_{m}$. For integer values of $m$ we have the simplification,
\begin{equation}
	\left(a+b\right)_{m}\left(a-b\right)_{m} = \prod_{k=0}^{m-1}\biggl\{\left(a+k\right)^{2}-b^{2}\biggr\}.
\end{equation}
Although $b$ in the above representations contains square root, in the final expressions we get polynomials in the Mandelstam variables. 
For $\alpha=0$ and $\beta=1$ we obtain \eqref{clst}, after using $(1-n-a)_{n-1}=(-1)^{n-1}(1+a)_{n-1}$.

\color{black}
\section{Low energy expansion and mathematical curiosities}
Let us expand \eqref{op-st} in terms of $x=s_1+s_2, y=s_1 s_2$. The Gamma function form after removing the massless pole, at $y=0$ becomes $-\sum_{k=0}^\infty \zeta(k+2) x^k+O(y)$. We expand the {\it rhs} and compare powers. This gives:
\begin{eqnarray}
\zeta(j)=\sum_{n=1}^\infty \sum_{k=0}^{n-1}\frac{(-1)^{k+n+1}}{n^2 n!}\left(\frac{\lambda}{n+\lambda}\right)^{j-2} \left(1-\frac{n\lambda}{n+\lambda}\right)^{k-j+2}
 \left( \frac{n(n+2\lambda)}{n+\lambda}{}^kC_{j-2}
+
\frac{n+\lambda-n\lambda}{\lambda}P^{(\alpha_k,\beta_k)}_{j-3}(\xi_n)\right)S_{n-1}^{(k)}\,,\nonumber
\end{eqnarray}
where $\xi_n=2\frac{n\lambda}{n+\lambda}-1$ and $P^{(\alpha,\beta)}(x)$ is the Jacobi Polynomial. Here $\alpha_k=3-j+k, \beta_k=-k$. $S_{n-1}^{(k)}$ are the Stirling numbers of the first kind and ${}^k C_j$ is the binomial coefficient.
A generating function for Zeta function of the form $Z(x)\equiv\sum_{k=2}^\infty \zeta(k) x^k=-x H_{-x}$ in terms of Harmonic number $H_{-x}$, follows from the low energy expansion and is given by
\begin{eqnarray}
	Z(x)
	&=&\sum_{n=1}^\infty \frac{x^2}{n!} \left(\frac{2n-x}{n^2-n x}-\frac{1}{\lambda+n}\right)
	 \left(1+\frac{\lambda(x-n)}{\lambda+n}\right)_{n-1}. \nonumber
\end{eqnarray}
Finally, we present a formula for $\pi$ which can be obtained by setting $s_1=-1/2=s_2$ in the open string amplitude,
\begin{eqnarray}
	\pi=4+\sum_{n=1}^\infty \frac{1}{n!}\left(\frac{1}{n+\lambda}-\frac{4}{2n+1}\right)\left(\frac{(2n+1)^2}{4(n+\lambda)}-n\right)_{n-1}.\nonumber
\end{eqnarray}
In the $\lambda\rightarrow \infty$ limit, the summand goes over to $(-1)^{n} 4/(2n+1)$ which is precisely the Madhava series! While this series takes 5 billion terms to converge to 10 decimal places, the new representation with $\lambda$ between $10-100$ takes 30 terms.



\section{Beta-function series representations}
In the main text we presented the standard representation
\be\label{repsupp1}
B(-s_1,-s_2)=-\frac{1}{s_1}-\sum_{n=1}^\infty \frac{1}{n!} \frac{(s_2+1)_n}{s_1-n}\,,
\ee
which converges for ${\rm Re}(s_2)<0$ and a similar one with $s_1, s_2$ interchanged.
Naively, one may think that to get a representation that manifests poles in both variables, we simply take 
\be\label{rep1}
\frac{1}{2}\left(-\frac{1}{s_1}-\sum_{n=1}^\infty \frac{1}{n!} \frac{(s_2+1)_n}{s_1-n}\right)+\frac{1}{2}\left(-\frac{1}{s_2}-\sum_{n=1}^\infty \frac{1}{n!} \frac{(s_1+1)_n}{s_2-n}\right)\,.
\ee
However, this only converges when ${\rm Re}(s_1,s_2)<0$ and is unsuitable for truncation and consideration as something that would arise in a QFT. In fact, it is obvious that due to the factor of $1/2$ we will get the wrong residues in both channels and the reason is simply because the representation does not converge in the  ${\rm Re}(s_1,s_2)>0$, where the poles lie. Recall that for a representation to be considered suitable for a QFT interpretation, it should not have any non-analyticity except at the poles corresponding to the mass-levels. To drive home this point further, we will present a comparison of various possibilities in the table below. But before making this comparison, let us consider two other interesting possiblities.

In addition to the standard representation for the Euler-Beta function in the main text, it also has the integral representation \cite{gradshteyn2014table}:
\be\label{rep2}
B(-s_1,-s_2)=\int_0^1 dz\, \frac{z^{-s_1-1}+z^{-s_2-1}}{(1+z)^{-s_1-s_2}}\,,
\ee
which converges when ${\rm Re}(s_1),{\rm Re}(s_2)<0$. If we expand the denominator around $z=0$ and integrate term by term, we will get an analytically continued representation in terms of the sum over poles in both $s_1, s_2$, albeit one which only works in a limited regime:
\be\label{rrep2}
\sum_{n=0}^\infty \left(\frac{1}{n-s_1}+\frac{1}{n-s_2}\right)\frac{(s_1+s_2-n+1)_n}{n!}\,,
\ee
where the summand for large $n$ goes like $n^{-2-s_1-s_2}$ needing ${\rm Re}(s_1+s_2)>-1$ for convergence.
Contrary to what is frequently mentioned in the older literature, this form, which sums over poles in both channels does not have any ``double-counting''---the crucial, implicit contact terms in this representation avoid this. However, the restriction  ${\rm Re}(s_1+s_2)>-1$ is undesirable since in tree-level QFT, we do not expect any further singularities in the complex $s_1,s_2$-plane except at the poles {\it in all channels}. Thus, while promising, this representation is still not what we would identify with QFT. Note, however, that the summand in the new representation in the main text, for $n$ fixed, in the $\lambda\gg 1$ limit, becomes exactly the summand in the above representation. 

In our opinion, so far, the most interesting possibility in the literature arises using the results of Giddings from 1986, using Witten's open string field theory. The upshot of the analysis is that the $s_1$ and $s_2$ channels can be identified by splitting the defining integral into two pieces:
\be
\int_{0}^{1/2} dz \,z^{-s_1-1}(1-z)^{-s_2-1}+\int_{1/2}^{1} dz \,z^{-s_1-1}(1-z)^{-s_2-1}\,.
\ee
While it does not seem to have been considered before, we can
simply expand the first integrand around $z=0$ and the second around $z=1$, and integrate term by term and obtain the (analytically continued) representation:
\be\label{repGiddings}
\sum_{n=0}^\infty \frac{(-\frac{1}{2})^n}{n!}\left(\frac{2^{s_1}(-n-s_2)_n}{n-s_1} +\frac{2^{s_2}(-n-s_1)_n}{n-s_2}\right)\,. 
\ee
The fact that it is the  analytic continuation of the integral form can be checked by matching with it for ${\rm Re}(s_1,s_2)<0$. This has the correct analytic structure and has poles in both channels and works reasonably well in some regimes of $s_1, s_2$ as the table above shows. 
However, even this does not meet all the criteria to be called a QFT representation. For starters, the $2^{s_1}$ factor means that for each mode, we need to expand around the pole and get an infinite number of polynomials---one will need to find a suitable prescription how to truncate this. Also the truncation does not depend on any parameter. However, even otherwise, it is clear that for $s_1, s_2>0$, convergence is going to be poor. This is indeed verified in the table, which demonstrates that none of the representations above compare to the accuracy and usefulness of the new representation presented in the main text.

\color{magenta}
\begin{center}
	\begin{table}
		\begin{tabular}{|c|c|c|c|c|c|}
			\hline
			Parameters & Actual answer & \begin{tabular}{c}
				Eq.(\ref{rep1})\\(No. of terms = 25)
			\end{tabular} & \begin{tabular}{c}
				Eq.(\ref{rrep2})\\ (No. of terms = 25)
			\end{tabular}  & \begin{tabular}{c}
				Eq.(\ref{repGiddings}) \\ (No. of terms = 25)
			\end{tabular}& \begin{tabular}{c}
				New representation\\ (No. of terms = 25)\\ $\lambda= 8.5$
			\end{tabular}\\
			\hline
			&&&&&\\
			$s_{1}=17.5$, $s_{2}=-15.2$  & -0.00021590 & $1.1961\times10^{10}$ & -0.00021454 & -0.028595 & -0.00021590\\
			&&&&&\\
			$s_{1} = 7.1$, $s_{2}= -1.1$  & 0 & $3.4765\times10^{5}$ & 0 & -0.0032111 & $9.9293\times10^{-8}$\\
			&&&&&\\
			$s_{1}=-7.6$, $s_{2}=0.5$  & -9.2810 & -2.1050 & -33543 & -9.2810 & -9.2810\\
			&&&&&\\
			\hline
		\end{tabular}
		\caption{We compare various representations.}
	\end{table}
\end{center}

\normalcolor
\section{Open-string amplitude: Contact terms}
The Pochhammer symbol appearing in the open string amplitude representation can be written as
\begin{equation}
	\left(1-\lambda+\frac{\left(s_{1}+\lambda\right)\left(s_{2}+\lambda\right)}{\lambda+n}\right)_{n-1} = \prod_{k=1}^{n-1}\left[k-\lambda+\frac{\left(s_{1}+\lambda\right)\left(s_{2}+\lambda\right)}{\lambda+n}\right].
\end{equation}
Using this we can write down the contact terms, $\kappa_{n}$ at $n$-th level as follows,
\begin{eqnarray}
	\kappa_{n} & = & \frac{1}{n!\left(s_{1}-n\right)}\left(1+s_{2}\right)_{n-1}\Biggl\{\prod_{k=1}^{n-1}\left[1+\frac{\left(s_{1}-n\right)\left(s_{2}+\lambda\right)}{\left(k+s_{2}\right)\left(n+\lambda\right)}\right]-1\Biggr\}+ \frac{1}{n!\left(s_{2}-n\right)}\left(1+s_{1}\right)_{n-1}\Biggl\{\prod_{k=1}^{n-1}\left[1+\frac{\left(s_{2}-n\right)\left(s_{1}+\lambda\right)}{\left(k+s_{1}\right)\left(n+\lambda\right)}\right]-1\Biggr\}\nonumber\\
	&& + \frac{1}{n!\left(\lambda+n\right)}\left(1-\lambda+\frac{\left(s_{1}+\lambda\right)\left(s_{2}+\lambda\right)}{\lambda+n}\right)_{n-1}.
\end{eqnarray}
The terms are regular in $s_1,s_2$ as can be seen. An explicit formula, involving Stirling numbers of the first kind, can also be worked out for the same using the formula (and its $s_1\leftrightarrow s_2$ version) below:
\be
\left(1-\lambda+\frac{\left(s_{1}+\lambda\right)\left(s_{2}+\lambda\right)}{\lambda+n}\right)_{n-1}=(1+s_2)_{n-1}+\sum_{k=0}^{n-1}\sum_{j=1}^{k}(-1)^{k+n-1}S^{(k)}_{n-1}\,{}^k C_j \left[\frac{(s_1-n)(s_2+\lambda)}{n+\lambda}\right]^j (1+s_2)^{k-j}\,. 
\ee
The first term on the {\it rhs} gives the residue for the $s_1=n$-pole, while the next one leads to the regular contact terms.

\section{Checks of the new representation}
\begin{figure}[H]
	\centering
	\begin{subfigure}{0.5\textwidth}
			\includegraphics[scale=0.5]{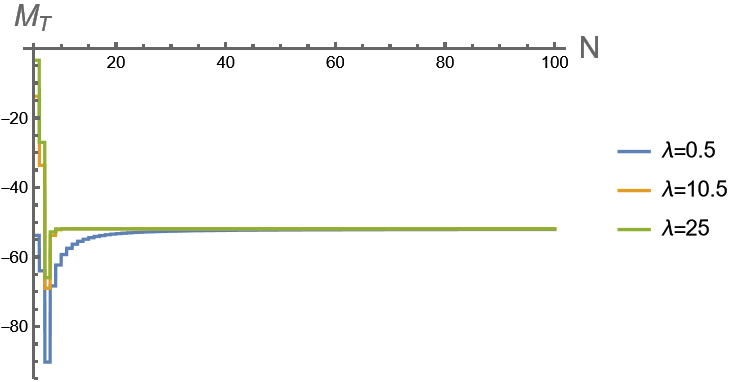}
		\subcaption{Truncated sum, $M_{T}\left(s_{1}, s_{2}\right)$ plotted against $N$, number of terms in the sum for different values of $\lambda$. Here $s_{1}=7.5$ and $s_{2}=1.2$.} \label{MTvN}
	\end{subfigure}
\hspace{0.5cm}
	\begin{subfigure}{0.45\textwidth}
	\includegraphics[scale=0.5]{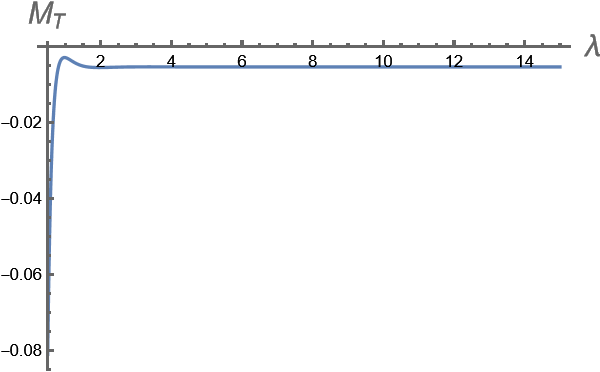}
	\subcaption{$M_{T}\left(s_{1},s_{2}\right)$ with $N=20$ plotted against $\lambda$. Here $s_{1}=7.5$ and $s_{2}=-3.2$.}\label{MTvlam}
\end{subfigure}
\caption{}
\end{figure}

In fig.(\ref{MTvN}) we show that different choices of $\lambda$ give the same answers as $N\rightarrow \infty$. This is a non-trivial check for the claims in the main text. 
In fig.(\ref{MTvlam}) we give another example for the plateau which enables us to read off the ``true'' value of the amplitude for the truncated representation.

\begin{figure}[H]
	\centering
	\includegraphics[scale=0.8]{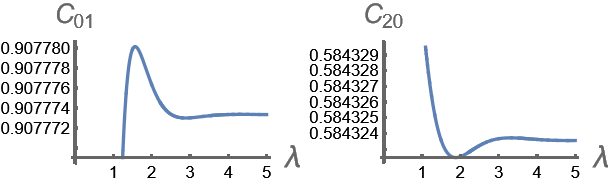}
	\caption{Coefficients of contact terms appearing at $\mathcal{O}\left(s_{i}^{2}\right)$ plotted against $\lambda$. Sum is truncated at $N=20$}.\label{contactfig}
\end{figure}
In fig.(\ref{contactfig}) we show the contact terms as a function of $\lambda$.
We denote the contact terms by $\left(-1\right)^{q}C_{pq}(s_1+s_2)^{p}(s_1 s_2)^{q}$.  We show in fig.(\ref{contactfig}) how the coefficients $C_{pq}$ change with $\lambda$. A distinct plateau appears with $\lambda$. These plateaus indicate that $\partial_\lambda M_T\approx 0$, which is exactly what we would want in the QFT representation.

\section{Locality/Null constraints}
We recall the two-channel symmetric dispersion relation \cite{Raman:2021pkf} which is given by 
\begin{eqnarray}\label{disp-2ch-s}
	\mathcal{M}\left(s_{1},s_{2}\right) & = &  \mathcal{M}\left(0,0\right) + \frac{1}{\pi}\int_{a}^{\infty}\frac{\mathrm{d}\sigma}{\sigma}\mathcal{A}(\sigma,\frac{a \sigma}{\sigma-a})\left[\frac{s_{1}}{\sigma-s_{1}}+\frac{s_{2}}{\sigma-s_{2}}\right]. 
\end{eqnarray}
Using the above equation we can obtain another representation for open string amplitude,
\begin{eqnarray}\label{opst-nl}
	\frac{\Gamma\left(-s_{1}\right)\Gamma\left(-s_{2}\right)}{\Gamma\left(1-s_{1}-s_{2}\right)} & = & \frac{\Gamma^{2}\left(\lambda\right)}{\Gamma\left(1+2\lambda\right)} + \sum_{n=0}^{\infty}\frac{1}{n!}\left(\frac{1}{s_{1}-n}+\frac{1}{s_{2}-n}+\frac{2}{\lambda + n}\right)\left(1-\lambda+\frac{\left(s_{1}+\lambda\right)\left(s_{2}+\lambda\right)}{\lambda+n-\frac{\left(s_{1}-n\right)\left(s_{2}-n\right)}{\lambda+n}}\right)_{n-1}.
\end{eqnarray}
Here $\lambda$ is a free parameter and the \textit{lhs} of the above equation is independent of $\lambda$. For convergence of the above series we require ${\rm Re}(\frac{s_{1}s_{2}-\lambda^{2}}{s_{1}+s_{2}+2\lambda})<2$. The \textit{rhs} of \eqref{opst-nl} contains negative powers in $s_{1}$ and $s_{2}$ which come from the Pochhammer. Negative powers in Mandelstam variables usually signify non-local interactions in a theory. However, the other representation of the open string amplitude that we have provided in this paper,
\begin{eqnarray}\label{op-st2}
	\frac{\Gamma\left(-s_{1}\right)\Gamma\left(-s_{2}\right)}{\Gamma\left(1-s_{1}-s_{2}\right)} & = & \sum_{n=0}^{\infty}\frac{1}{n!}\left(\frac{1}{s_{1}-n}+\frac{1}{s_{2}-n}+\frac{1}{\lambda+n}\right)\left(1-\lambda+\frac{\left(s_{1}+\lambda\right)\left(s_{2}+\lambda\right)}{\lambda+n}\right)_{n-1}
\end{eqnarray}
does not have such non-local terms. Therefore, the non-local terms must eventually drop out from \eqref{opst-nl}. This leads us to null constraint sum rules, $\Delta$ which can be obtained by taking difference of the two representations given in equations (\ref{op-st2}) and (\ref{opst-nl}).

\begin{figure}[H]
	\centering
	\begin{subfigure}{0.3\textwidth}
		\includegraphics[scale=0.5]{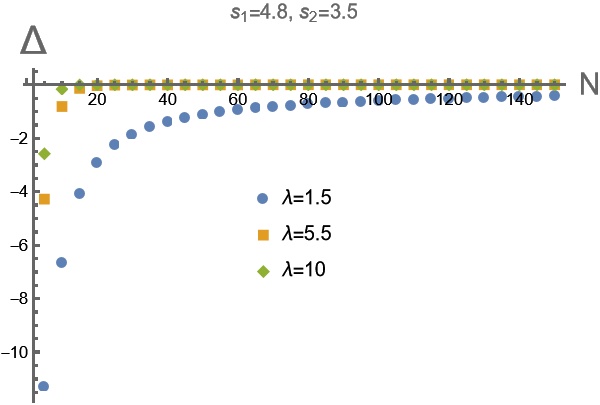}
	\end{subfigure}
	\qquad
	\begin{subfigure}{0.3\textwidth}
		\includegraphics[scale=0.5]{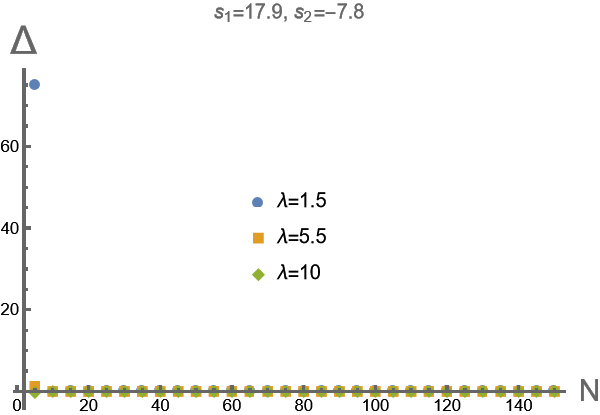}
	\end{subfigure}
	\qquad
	\begin{subfigure}{0.3\textwidth}
		\includegraphics[scale=0.5]{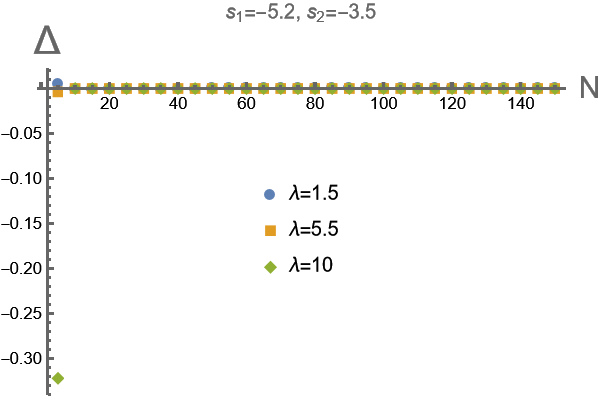}
	\end{subfigure}
	\caption{Plots of $\Delta= \text{Eq.(\ref{op-st2})} - \text{Eq.(\ref{opst-nl})}$ as number of terms in the sum, $N$ is varied for different choices of $s_{1}$, $s_{2}$ and $\lambda$. $\Delta$ diminishes as $N$ is increased. }
\end{figure}
Numerically as the number of terms in the series is made larger, $\Delta$ tends to zero. This gives numerical evidence that the null constraints are satisfied. 
\section{Tachyon scattering and other amplitudes}
The Euler-Beta function appears in the Veneziano amplitude which is the scattering of 4 open string tachyons. In this section, we would like to understand this amplitude using our QFT insights. The amplitude is fully crossing symmetric and is given by
\begin{equation}
	M(s,t)=B(-\alpha(s),-\alpha(t))+B(-\alpha(s),-\alpha(u))+B(-\alpha(t),-\alpha(u))\,,
\end{equation}
where $\alpha(s)=\alpha_0+\alpha' s$ and $s+t+u=4 m^2$. We will work with $\alpha'=1$. For tachyon scattering $\alpha_0=1$ and $m^2=-1$, but we will keep the discussion a bit more general, since the Euler-Beta function also arises in the description of scattering of other massive states. Let us focus on $B(-\alpha(s),-\alpha(t))$:
\be
B(-\alpha(s),-\alpha(t))=\frac{\Gamma(-\alpha_0-s)\Gamma(-\alpha_0-t)}{\Gamma(-2\alpha_0-s-t)}\,.
\ee
In the notation of the generalized Euler-Beta function used in the main text, we have $\alpha=-\alpha_0, \beta=-2\alpha_0$, $s_1=s, s_2=t$. Thus $2\alpha-\beta=0$, which meets the criteria for having polynomial residues. For the convergence of the polynomial contact terms, we want $\alpha<\beta$. This is satisfied if $\alpha_0<0$, which is the situation when we want to describe scattering of positive mass string states. However, for tachyon scattering the situation is different since $\alpha_0=1>0$. Since $s,t$ are the usual Mandelstam variables, and in examining the convergence of the contact terms, we have expanded around $s,t\sim 0$, this is a legitimate thing to do when we have scattering of massless or positive mass particles. However, the tachyon is a negative mass-squared particle and expanding around $s,t\sim 0$ is not sensible as from the perspective of the tachyon this is high-energy! Rather, we present the contact terms by expanding around another point. Let us shift the variables to $s=s_1+\gamma, t=s_2+\gamma$. Then for convergence of the contact terms which are now polynomials in $s_1,s_2$, we would need $\gamma<-\alpha_0$. To maintain full 3-channel crossing symmetry, we must choose $\gamma=-4/3$. Now, we have a sensible QFT expansion of the tachyon amplitude.

\end{document}